\documentclass[12pt]{article}
\usepackage{epsfig}
\newcommand{\preprint}[1]{\rule{0pt}{8pt} \scriptsize #1}
\begin{document}

\parindent = 12pt

\pagestyle{empty}

\title{A Closer Look at two $AdS_4$ Branes in an $AdS_5$ Bulk}

\author{Shiyamala Thambyahpillai\footnote{thamby@physics.harvard.edu}}
\maketitle
\begin{picture}(0,0)
\put(400,200){\shortstack{
        \preprint HUTP-04/A035\\
        \preprint hep-th/0409190\\
        \rule{0pt}{8pt} }}
\end{picture}
\begin{abstract}
\noindent
We investigate a scenario with two $AdS_4$ branes in an $AdS_5$ bulk. In this scenario  there are two gravitons and we investigate the role played by each of them for different positions of the second brane. We show that both gravitons play a significant role only when the turn-around point in the warp factor is approximately equidistant from both branes. We find that the ultralight mode becomes heavy as the second brane approaches the turn-around point, and the physics begins to resemble that of the RS model. Thus we demonstrate the crucial role played by the turn-around in the warp factor in enabling the presence of both gravitons.
\end{abstract}

\section{Introduction}
There has been much interest recently in brane world scenarios and their implications for large extra dimensions. Whilst the SM fields are confined to the brane, gravity propogates in the 5D bulk. In conventional models, the Newton potential between two masses on the brane results from the exchange of a massless graviton and a tower of massive KK modes. However, the model proposed by Karch and Randall \cite{kr} gives the first instance where 4D gravity is mediated by a massive (though light) graviton, and there is no massless mode. In this scenario the brane has a small negative 4D cosmological constant on it. The warp factor blows up far from the brane, and the geometry includes the boundary of $AdS_5$ and thus the massless mode is nonnormalizable. However the first KK mode is ultralight in comparison to the rest of the tower, and it is this mode which now plays the role of the graviton.
\\ \\
The introduction of a second brane to cut off the extra dimension means the normalization of the massless mode is made finite. We are then left with a situation where we have two gravitons - a massless mode and an ultralight mode (we are assuming a small 4D cosmological constant). This model was investigated by Kogan et al \cite{big} in the case that the two branes are equidistant from the turn-around point in the warp factor. The concept of having two gravitons with interchanging roles is fascinating, and merits further investigation, which is the subject of this paper.
\\ \\
We investigate what happens to the two gravitons when the second brane is placed at an arbitrary postion, then we look at three specific limits in order to simplify matters. We look at whether the massless and ultralight modes couple with equal strength to the branes, and where they are concentrated in these limits. We are particularly interested in what happens when the second brane approaches the turn-around point in the warp factor. In this case we find the mass of the ultralight mode blows up and we are left with only the massless graviton. This presents a scenario similar to the RS model, demonstrating the necessity of the kink in the warp factor to producing a situation involving two gravitons (or more specifically an ultralight mode). Conversely when the second brane is taken to infinity, the massless mode is no longer normalizable and we are left with only the ultralight mode as in \cite{kr}. We also calculate the Newton potential.
We investigate the origins of having two gravitons in the theory by seeing what happens when we bring two $AdS_4$ branes from infinity to a finite separation. Finally we look at what happens to the two gravtions in the Minkowski limit, and how they combine to form the massless mode of the Randall-Sundrum model.

\section{The Background}
\noindent
We have 2 branes at positions $y = 0,y_c$. The action for this situation is:
\begin{equation}
S = \int d^4x \int_{-y_c}^{y_c} dy\sqrt{-g}( -\Lambda + 2M^3R ) +
\sum_{i}\int_{y=L_i} d^4x V_i \sqrt{\hat{g}^{(i)}}
\end{equation}

\noindent
where $\hat{g}^{(i)}$ is the induced metric on the branes; $\Lambda, M$ are
the 5-d cosmological constant and fundamental gravity scale respectively.
\\ \\
The metric is as follows:

\begin{eqnarray}
&&ds^2 = e^{2A(y)}g_{ij}dx^idx^j + dy^2
\nonumber\\
&&e^{A(y)} = \frac{\cosh (\frac{y_0 - |y|}{L})}{\cosh (\frac{y_0}{L})}
\end{eqnarray}

\noindent
where $y_0$ is the turn-around point in the warp factor and $\frac{1}{L} = \sqrt{\frac
{-\Lambda}{24 M^3}}$. The warp factor has been normalized to one at the position of the first brane (at 0).
\\ \\
The brane tensions depend on the derivative of the warp factor at the position of each brane. They are:
\begin{eqnarray}
&& V_1 = \frac{3}{L}\tanh (\frac{y_0}{L})
\nonumber\\
&& V_2 = \frac{3}{L}\tanh (\frac{y_c-y_0}{L})
\label{eqn:V}
\end{eqnarray}

We can find the 4-d cosmolgical constant on each brane i:
\begin{eqnarray}
&& -\Lambda_4 \sim \frac{1}{L^2}\frac{1}{\cosh^2 (\frac{y_i-y_0}{L})}
\nonumber\\
&&= \frac{1}{L^2} (1-(\frac{L}{3})^2 V_i^2)
\end{eqnarray}

We wish for the 4-d cosmological constant on our brane to be small, as this is essential if we want the first KK mode to be ultralight. If we live on the brane at $y=0$ this corresponds to large $\frac{y_0}{L}$, which we assume from now on. Then the brane tension $V_{1}\rightarrow 3/L$.
\\ \\
We will be using the conformal coordinate $z$ which is related to $y$ by
\begin{eqnarray}
&&\tan (\frac{1}{2\hat{L}}(|z| - z_0 )) \equiv \tanh (\frac{1}{2L}(|y| - y_0 ))
\nonumber\\
&&\frac{1}{\hat{L}} = \frac{1}{L\cosh (\frac{y_0}{L})} = \frac{\cos (\frac{z_0}{\hat{L}})}{L}
\label{eq:z}
\end{eqnarray}

\section{The 2 Gravitons}
\subsection{The Massless Mode}
\noindent
We start by looking at the role of the massless mode. Its wavefunction is

\begin{equation}
\hat{\psi}^{(0)}(z) = \frac{A}{\cos^{\frac{3}{2}} (\frac{(|z| - z_0 )}{\hat{L}})}
\label{eq:0mode}
\end{equation}

\noindent
where the normalization constant $A$ can be obtained from:
\\ \\
$\int_{-z_c}^{z_c}  dz \hat{\psi}^2 = 1$
\\ \\
so
\begin{equation}
A^2 = \frac{1}{\hat{L}\Biggl(
\frac{\sin ((z_c - z_0 )/\hat{L})}
{\cos^2 ((z_c - z_0 )/\hat{L})} + \frac{\sin (z_0/\hat{L})}
{\cos^2 (z_0/\hat{L})} +
\ln\tan \Biggl(\frac{1}{2\hat{L}}(z_c -z_0) + \frac{\pi}{4}\Biggr)
+ \ln\tan \Biggl(\frac{1}{2\hat{L}} (z_0 ) + \frac{\pi}{4}\Biggr)\Biggr)}
\label{eq:norm}
\end{equation}

\noindent
Let us consider three specific values for the position of the second brane,
which are sufficient to give us an idea of what is happening in general.\\
We investigate what happens when the second brane is sent to infinity, when it
is at $2z_0$ (symmetric configuration) and in the limit that it approaches
 the turn-around in the warp factor, $z_0$.

\subsubsection{\boldmath$z_{c} = z_{\infty}$}
\unboldmath
\noindent
First we send the 2nd brane to infinity, that is to a position $z_
{\infty}$ in conformal coordinates, given by:
\\ \\
$\frac{(z_{\infty}-z_0 )}{\hat{L}} = \frac{\pi}{2}$
\\ \\
then
\begin{equation}
A^2 \sim \frac{1}{\hat{L}\Biggl(\frac{1}{\cos^2 (\frac{\pi}{2})} + \ln\tan (\frac{\pi}
{2})\Biggr)} \rightarrow 0
\end{equation}

\noindent
We see the massless mode is not normalizable, it does not exist.\\
What happens to its wavefunction?
\begin{equation}
\hat{\psi}^{(0)}(z)^2 = \frac{A^2}{\cos^3 ((|z|-z_0)/\hat{L})} \rightarrow 0
\end{equation}

\noindent
for all $z$ except at $z_c =z_{\infty}$ where we get
\begin{equation}
\frac{1}{\cos^3 (\frac{\pi}{2})} \frac{1}{\frac{1}{\cos^2 (\frac{\pi}{2})}
+ ...}
 \rightarrow \infty
\end{equation}

\noindent
thus the mode is concentrated at the 2nd brane at infinity.

\subsubsection{\boldmath$z_{c} = 2z_0$}
\unboldmath
\noindent
 What happens if the branes are placed in a symmetric configuration around the
turn-around in the warpfactor, as in \cite{big} ($z_{c} = 2z_0$)?\\
then
\begin{eqnarray}
&&\hat{\psi}^{(0)}(z )^2 = \frac{\cos (z_0/\hat{L} )}
{2L\cos^3 ((|z|-z_0 )/\hat{L})} \frac{1}{\frac{\sin (z_0/\hat{L})}
{\cos^2 (z_0/\hat{L})} + \ln\tan (\frac{z_0}{2\hat{L}} + \frac{\pi}{4})}
\nonumber\\
&&\cos (z_0/\hat{L}) \equiv \frac{1}{\cosh (y_0/L)} \approx 2e^{-y_0/L}
\end{eqnarray}
\\ for large $y_0/L$, so:
\begin{equation}
\hat{\psi}^{(0)}(z)^2 \approx \frac{1}{2L}\frac{\cos^3 (z_0/\hat{L})}
{\cos^3 ((|z| - z_0)/\hat{L})}
\end{equation}

\noindent
thus
\begin{equation}
\hat{\psi}^{(0)} (z_c = 2z_0 ) = \hat{\psi}^{(0)} (0) = \frac{1}{2L}
\label{eq:hmmm}
\end{equation}

\noindent
The massless mode is equally distributed on both branes as expected.

\subsubsection{\boldmath$z_{c} = z_0$}
\unboldmath
\noindent
What happens if the 2nd brane actually approaches the turn-around in the
warpfactor?

\begin{eqnarray}
\hat{\psi}^{(0)}(z)^2 &=& \frac{\cos (z_0/\hat{L})}{L\cos^3 ((|z|-z_0)/\hat{L})} \frac{1}{\frac{\sin (z_0/\hat{L})}
{\cos^2 (z_0/\hat{L})} + \ln\tan (\frac{z_0}{2\hat{L}}+\frac{\pi}{4})}
\nonumber\\
&&\approx \frac{\cos^3 (z_0/\hat{L})}{L\cos^3 ((|z|-z_0)/\hat{L})}
\end{eqnarray}

\noindent
so $\hat{\psi}^{(0)} (0) = \frac{1}{L}$,   $\hat{\psi}^{(0)} (z_c =z_0 )
\sim\frac{8}{L}e^{-\frac{3}{L}y_0}$   which is very small.
The massless mode is now concentrated on the first (Planck) brane. For comparison, in RS1 the massless mode is also concentrated on the Planck brane.

\subsection{The Ultralight 1st KK Mode}

\noindent
Now that we have studied the massless mode in some detail, let us look at the
 ultralight 1st KK mode, which plays the role of the 2nd graviton in this
scenario.
\\ \\
The KK states obey:

\begin{eqnarray}
[\partial_{z}^2 &-& V(z)]\hat{\psi}^{(n)} (z) = -m^2\hat{\psi}^{(n)} (z)
\nonumber\\
\nonumber\\
V(z) &=& -\frac{9}{4\hat{L}^2} + \frac{15}{4\hat{L}^2}\frac{1}
{\cos^2 ((|z|-z_0)/\hat{L})}
\nonumber\\
&-& \frac{3}{\hat{L}}
[\tan (z_0/\hat{L})\delta (z) +
\tan ((z_c -z_0 )/\hat{L})\delta (z-z_c )]
\label{eq:potnkk}
\end{eqnarray}
\\ \\
\\ \\
\\ \\
\\ \\
\noindent
which can be shown to have the form of a hypergeometric differential equation
which has the following solution:

\begin{eqnarray}
\hat{\psi}^{(n)} &=& \cos^{\frac{5}{2}} ((|z| - z_0 )/\hat{L})
\Biggl[C_1 F_{\frac{5}{4}} - C_2 \sin((|z| -z_0)/\hat{L})
F_{\frac{7}{4}}\Biggr]
\nonumber\\
\nonumber\\
F_{\frac{5}{4}} &=& F\Biggl(\frac{5}{4} + \frac{1}{2}
\sqrt{\Biggl(\frac{m}{\hat{k}}\Biggr)^2 + \frac{9}{4}},
\frac{5}{4} - \frac{1}{2}
\sqrt{\Biggl(\frac{m}{\hat{k}}\Biggr)^2 + \frac{9}{4}}, \frac{1}{2},
\sin^2 ((|z|-z_0)/\hat{L})\Biggr)
\nonumber\\
&=&F(\alpha ,\beta ,\gamma ,\sin^2 ((|z|-z_0)/\hat{L}))
\nonumber\\
\nonumber\\
F_{\frac{7}{4}} &=& F\Biggl(\frac{7}{4} + \frac{1}{2}
\sqrt{\Biggl(\frac{m}{\hat{k}}\Biggr)^2 + \frac{9}{4}},
\frac{7}{4} - \frac{1}{2}
\sqrt{\Biggl(\frac{m}{\hat{k}}\Biggr)^2 + \frac{9}{4}}, \frac{3}{2},
\sin^2 ((|z|-z_0 )/\hat{L})\Biggr)
\nonumber\\
&=&F(\alpha' ,\beta' ,\gamma' ,\sin^2 ((|z| - z_0 )/\hat{L}))
\label{eq:hyper}
\end{eqnarray}

\noindent
$C_1, C_2$ are constants.
\\ \\
The following expansion for hypergeometric functions proves very useful,
especially for small $\cos^2 ((|z|- z_0 )/\hat{L}) = (1-x) =
\epsilon$:
\begin{eqnarray}
F(a,b&,&a+b-m,x) = \frac{\Gamma (m)\Gamma (a+b-m)}{\Gamma (a)\Gamma (b)}
(1-x)^{-m}\Biggl[\sum_{k=0}^{m-1}\frac{(a-m)_k (b-m)_k}{k!(1-m)_k}(1-x)^k
\Biggl]
\nonumber\\
&&-(-1)^m\frac{\Gamma (a+b-m)}{\Gamma (a-m)\Gamma (b-m)}
\Biggl[\sum_{k=0}^{\infty}\frac{(a)_k (b)_k}{k!(k+m)!}(1-x)^k
\nonumber\\
&&.[\ln (1-x) - \psi (k+1)-\psi(k+m+1)+\psi(a+k)+\psi(b+k)]\Biggr]
\label{eq:prudnikov}
\end{eqnarray}

\noindent
where $m$ is an integer, and
\\
$(a)_k=a(a+1)...(a+k-1)$ ,  $ (a)_0=1$

\subsubsection{Finding the Mass}
\noindent
Using the boundary conditions at the branes:
\begin{equation}
\hat{\psi}' - \frac{3}{2\hat{L}}\tan((|z|-z_0)/\hat{L})\hat{\psi}=0
\label{eq:bc}
\end{equation}

\noindent
we find the following constraint
\begin{eqnarray}
C_1\Biggl[-4\tan (.) F_{\frac{5}{4}}&+&2\sin (.)\cos (.)\frac{\alpha\beta}
{\gamma}F_{\frac{9}{4}}\Biggr]
\nonumber\\
&&=C_2\sin (.)\Biggl[-4\tan (.) F_{\frac{7}{4}}+2\sin (.)\cos (.)
\frac{\alpha'\beta'}{\gamma'}F_{\frac{11}{4}}+\cot (.)F_{\frac{7}{4}}\Biggr]
\nonumber\\
\label{eq:bc2}
\end{eqnarray}

\noindent
where $\sin (.) \equiv \sin ((|z| - z_0 )/\hat{L})$
\\ \\
Let us relabel $\epsilon = \cos^2 ((|z| -z_0 )/\hat{L})$, and define 
$\epsilon_0$ as its value at the first brane and $\epsilon_c$ as its value at the 2nd
 brane. $\epsilon_0$ is small, and assuming that $\epsilon_c$ is also small
(which is certainly true for $z_c\geq 2z_0$ ) we find upon
substituting for the $F$'s using (\ref{eq:prudnikov})that:

\begin{eqnarray}
\Biggl(\frac{|F_{\frac{5}{4}}|}
{|F_{\frac{7}{4}}|}\Biggr)\frac{C_1}{C_2}&=&
\Biggl(\frac{\Gamma_{\frac{7}{4}+}\Gamma_{\frac{7}{4}-}}
{\Gamma_{\frac{5}{4}+}\Gamma_{\frac{5}{4}-}}\Biggr)\frac{2C_1}{C_2}
\nonumber\\
&=& -\Biggl(1-\frac{3\epsilon_0}{E}\Biggr)
\nonumber\\
&=& \Biggl(1-\frac{3\epsilon_c}{E}\Biggr)
\nonumber\\
\Rightarrow E=\frac{3}{2}(\epsilon_0+\epsilon_c)
\nonumber\\
\label{eq:emma}
\end{eqnarray}

\noindent
Where we have relabeled $(m\hat{L})^2$ as $E$ and are assuming it is
reasonably small ($<<1$) for the ``ultralight'' mode. Also we are using the following short-hand notation:
\begin{eqnarray}
&&|F_{\frac{5}{4}}|\equiv\frac{\Gamma (\frac{1}{2})}{\Gamma_{\frac{5}{4}+}\Gamma_{\frac{5}{4}-}}
\nonumber\\
&&|F_{\frac{7}{4}}|\equiv\frac{\frac{1}{2}\Gamma (\frac{1}{2})}{\Gamma_{\frac{7}{4}+}\Gamma_{\frac{7}{4}-}}
\end{eqnarray}

\noindent
where

\begin{eqnarray}
&&\Gamma_{\frac{7}{4}\pm} = \Gamma (\frac{7}{4}\pm\frac{1}{2}\sqrt{E+\frac{9}
{4}})
\nonumber\\
&&\Gamma_{\frac{5}{4}\pm} = \Gamma (\frac{5}{4}\pm\frac{1}{2}\sqrt{E+\frac{9}
{4}})
\label{eq:gamma}
\end{eqnarray}

\noindent
Substituting back for $E$ we find the coefficients are related by
\begin{equation}
\Biggl(\frac{|F_{\frac{5}{4}}|}{|F_{\frac{7}{4}}|}\Biggr)\frac{C_1}{C_2}
= \frac{\epsilon_0 -\epsilon_c}{\epsilon_0 +\epsilon_c}
\label{eq:coeff}
\end{equation}

\noindent
Where are the wavefunctions concentrated when the 2nd brane is at some
general position $z_c$?
Keeping the condition that $\epsilon_c$ is small, and expanding the $F$'s we
find
\begin{eqnarray}
\hat{\psi}^{(1)}(z)&=&\cos^{-\frac{3}{2}}\Biggl[C_1\Biggl
(|F_{\frac{5}{4}}|(1+\frac{3}{8}\cos^2(.)(\epsilon_0 +\epsilon_c))\Biggr)
\nonumber\\
&&-C_2\sin(.)
\Biggl(|F_{\frac{7}{4}}|(1+\frac{\cos^2(.)}{2} +
\frac{3}{8}\cos^2(.)(\epsilon_0 +\epsilon_c))\Biggr)\Biggr]
\end{eqnarray}

\noindent
so at the branes
\begin{eqnarray}
&&\hat{\psi}^{(1)}(0)\sim\frac{2\epsilon_0^{\frac{1}{4}}}
{\epsilon_0 +\epsilon_c}
\nonumber\\
&&\hat{\psi}^{(1)}(z_c)\sim\frac{-2\epsilon_c^{\frac{1}{4}}}
{\epsilon_0 +\epsilon_c}
\label{eq:conc}
\end{eqnarray}

\noindent
Finally we can also calculate the normalization constants of the wavefunction,
provided we assume that $E$ is sufficiently small. Let $C_1=\alpha C_2$, then:

\begin{eqnarray}
\frac{1}{\hat{L}C_2^2}&=&\Biggl[\alpha^2 (|F_{\frac{5}{4}}|)^2\Biggl(
\frac{\sin (.)}{\cos^2(.)}+\ln\tan (\frac{\pi}{4}+
\frac{1}{2\hat{L}}(z-z_0))\Biggr)
\nonumber\\
&-&2\alpha(|F_{\frac{5}{4}}|)(|F_{\frac{7}{4}}|)\frac{1}{\cos^2(.)}
+(|F_{\frac{7}{4}}|)^2\Biggl(\frac{\sin (.)}{\cos^2(.)}-\ln\tan
(\frac{\pi}{4}+\frac{1}{2\hat{L}}(z-z_0))\Biggr)\Biggr]_0^{z_c}
\nonumber\\
\label{eq:norm1}
\end{eqnarray}

\noindent
Now let us see what all this means at specific positions of the 2nd brane.

\subsubsection{\boldmath$z_{c} = z_{\infty}$}
\unboldmath

\noindent
In this case $\epsilon_c\rightarrow 0$ and (\ref{eq:emma}) tells us that
$E=\frac{3}{2}\epsilon_0$. Then
\\ \\
$E=\Biggl(m\hat{L}\Biggr)^2=\Biggl(mL\Biggr)^2\cosh^2(\frac{y_0}{L})$
\\ \\
$\Rightarrow m=\frac{4}{L}\sqrt{\frac{3}{2}}e^{-\frac{2}{L}y_0}$
\\ \\
is the mass of this ultralight mode. Also
\\ \\
$\Biggl(\frac{|F_{\frac{5}{4}}|}{|F_{\frac{7}{4}}|}\Biggr)\frac{C_1}{C_2}
=1$
\\ \\
so
\\ \\
$\hat{\psi}^{(1)}(z)=\cos^{\frac{5}{2}}C_2\Biggl[\Biggl(
\frac{|F_{\frac{7}{4}}|}{|F_{\frac{5}{4}}|}\Biggr)F_{\frac{5}{4}}
-\sin (.)F_{\frac{7}{4}}\Biggr]$
\\ \\
now from (\ref{eq:conc}) we see that
\begin{eqnarray}
&&\hat{\psi}^{(1)}(0)\sim\ 2\epsilon_0^{-\frac{3}{4}}
\nonumber\\
&&\hat{\psi}^{(1)}(z_c)\sim\frac{-2\epsilon_c^{\frac{1}{4}}}{\epsilon_0}
\sim 0
\label{eq:concinfty}
\end{eqnarray}

\noindent
So we find that the ultralight mode is concentrated on the first brane as the
2nd is taken to infinity. Earlier we saw how the massless mode is concentrated
on the 2nd brane in this case, and how it is not normalizable.
The same is not true of the ultralight mode:

\begin{eqnarray}
\frac{1}{\hat{L}C_2^2}&=&(|F_{\frac{7}{4}}|)^2\Biggl(\frac{1}
{\cos^2(\frac{\pi}{2})}+\ln\tan (\frac{\pi}{2})-\frac{2}{\cos^2(\frac{\pi}{2})}
\nonumber\\
&&+\frac{1}{\cos^2(\frac{\pi}{2})}-\ln\tan (\frac{\pi}{2})+ finite \Biggr)
\end{eqnarray}

\noindent
There is no diveregnce so $C_2$ is finite. Thus the 1st KK state is
normalizable even when the 2nd brane is taken away.

\subsubsection{\boldmath$z_c = 2z_0$}
\unboldmath

\noindent
For the symmetric configuration we see that $\epsilon_0 = \epsilon_c =
\epsilon$ so $C_1 = 0$ and
\\ \\
$\hat{\psi}^{(1)}(z)=C_2\sin (.)F_{\frac{7}{4}}$
\\ \\
then
\\ \\
$E=3\epsilon\Rightarrow m=\frac{4}{L}\sqrt{3}e^{-\frac{2}{L}y_0}=\sqrt{2}m_{\infty}$
\\ \\
Also (\ref{eq:conc}) tells us the wavefunction is equally concentrated on the
two branes for the symmetric configuration.
\\ \\
What happens for a slightly asymmetric configuration?
\\ \\
If $z_c$ is slightly $>2z_0$ then $\epsilon_c<\epsilon_0\Rightarrow
m<m_{symm}$. For instance, in y coordinates if the second brane is at position $2y_{0} + \delta $ then the mass becomes $m = m_{0} (1-\delta /2L)$. Also note that $\frac{C_1}{C_2}$ is small and positive. This results in the
ultralight mode being concentrated more on the 1st brane.
\\ \\
If $z_c$ is slightly $<2z_0$ then $\epsilon_c>\epsilon_0\Rightarrow
m>m_{symm}$ and $\frac{C_1}{C_2}$ is small and negative. The mode is more
concentrated on the 2nd brane.
\\ \\
So we see the mass of the 1st KK state decreases as the 2nd brane moves
further away, all the while coupling more strongly to the 1st brane.
\\ \\
Finally looking at the normalization of the wavefunction for the
symmetric configuration, (\ref{eq:norm1}) shows there is no divergence.
Setting $\alpha=0$ we find

\begin{equation}
\hat{\psi}^{(1)}(0)^2=\hat{\psi}^{(1)}(2z_0)^2=
\frac{\sin^2(.)\cos^{-3}(.)}{\hat{L}\Biggl(2\frac{\sin (z_0/\hat{L})}
{\cos^2((z_0/\hat{L}))}+\ln\Biggl(\frac{\tan (\frac{\pi}{4}+
\frac{z_0}{2\hat{L}})}{\tan (\frac{\pi}{4}-\frac{z_0}{2\hat{L}})}
\Biggr)\Biggr)}
\end{equation}

\noindent
We earlier found the wavefunction for the massless mode and so we see that
(\ref{eq:hmmm})
\begin{equation}
\hat{\psi}^{(0)}(0)^2\approx\frac{1}{2\hat{L}}\frac{1}{\cos (z_0/\hat{L})}
\sim\frac{1}{2L}\approx\hat{\psi}^{(1)}(0)^2
\label{eq:equal}
\end{equation}

\noindent
Thus the massless and ultralight gravitons have roughly the same couplings to
the branes in this special symmetric configuration.

\subsubsection{\boldmath$z_c = z_0$}
\unboldmath

\noindent
Perhaps the most interesting case occurs as the 2nd brane approaches the
turn-around in the warpfactor, $z_0$. In order to find the wavefunction of
 the 1st KK mode at the 2nd brane, let us use the relation:

\begin{equation}
F(a,b,c,x)=(1-x)^{(c-a-b)}F(c-a,c-b,c,x)
\label{eq:hyper2}
\end{equation}

\noindent
and then expand $F$ in terms of $x=\sin^2((z_c-z_0)/\hat{L})$ which is
 very small. We then substitute this into (\ref{eq:hyper}).
\\ \\
We need to find the boundary condition at $z_c$ and we do this by
plugging our value for $\hat{\psi}^{(1)}(z_c)$ into (\ref{eq:bc2}). We
find

\begin{eqnarray}
&&-EC_1=C_2 s_c( const-\frac{E}{2}+\frac{1}{s_c^2}+...)
\nonumber\\
&&\Rightarrow\frac{C_1}{C_2}\sim-\frac{1}{Es_c}
\nonumber\\
\nonumber\\
&&s_c\equiv\sin((z_c-z_0)/\hat{L})
\nonumber\\
\label{eq:Eturn}
\end{eqnarray}

\noindent
If we use (\ref{eq:Eturn}) together with the boundary condition from the brane
at $0$ (\ref{eq:emma}), we find that it is inconsistent to have $E$ small:

\begin{eqnarray}
\frac{C_1}{C_2}&=&-\frac{1}{Es_c}=
\Biggl(\frac{|F_{\frac{7}{4}}|}{|F_{\frac{5}{4}}|}\Biggr)
\Biggl(-1+\frac{3\epsilon_0}{E}\Biggr)
\nonumber\\
&&\Rightarrow-\frac{1}{s_c}\sim const(3\epsilon_0-E)
\nonumber\\
\end{eqnarray}

\noindent
This is impossible as $s_c$,$\epsilon_0$ and $E$ are all assumed to be very
small. Obviously this implies $E$ cannot be exactly zero either.
\\ \\
Is it possible that $E$ is not of order $\epsilon_0 $ but is still finite -
say
 $E<4$ ($E=4$ for the next KK mode)? We find from the boundary condition at $0$ (\ref{eq:emma})
 that

\begin{equation}
\frac{C_1}{C_2} = -const\equiv -\frac{1}{s_cE}
\end{equation}

\noindent
this is impossible as $s_c\rightarrow 0$
\\ \\
Say E is very large, or to be more precise
\begin{equation}
E\sim\frac{1}{s_c}\rightarrow\infty
\label{eq:trueE}
\end{equation}

\noindent
then the boundary condition at $z_c\rightarrow z_0$ (\ref{eq:Eturn}) gives us that
$\frac{C_1}{C_2}=-const$. A lengthy analysis of the boundary condition at $0$
in this situation indicates that
\\ \\
$\frac{C_1}{C_2}\Biggl(\frac{|F_{\frac{5}{4}}|}{|F_{\frac{7}{4}}|}\Biggr)
\sim-1$
\\ \\
Thus it is consistent to have very large $E\sim\frac{1}{s_c}$, which blows up
as the 2nd brane approaches the turn-around in the warp factor.
\\ \\
We can check that $E$ does indeed increase as we move the brane away from the
symmetric configuration at $z_c=2z_0$ towards $z_0$. Let us investigate how far the brane must be moved in order to change the order of magnitude of $E$ significantly. For instance, when is $E\sim\epsilon_0^{\frac{1}{2}}$? We find this occurs at
$\epsilon_c\sim\epsilon_0^{\frac{1}{2}}$, which is equivalent to $z_c=
1.95z_0$ if we let $y_0/L=5$. We can also show that $E\sim1$ at $z_c=
1.57z_0$
We find we lose the ``ultralight'' graviton very quickly.
\\ \\
Where is this 'not so light' 1st KK mode concentrated as $z_c\rightarrow
z_0$?
\\ \\
for $E\sim\epsilon_0^{\frac{1}{2}}$
\\ \\
$\hat{\psi}^{(1)}(0)\sim\Biggl(\frac
{\epsilon_0}{\epsilon_c}\Biggr)^{\frac{1}{4}}\epsilon_c^{-\frac{3}{4}}$
\\ \\
$\hat{\psi}^{(1)}(z_c)\sim\-\epsilon_c^{-\frac{3}{4}}$
\\ \\
and for $E\sim 1$
\\ \\
$\hat{\psi}^{(1)}(0)\sim\epsilon_0^{\frac{1}{4}}$
\\ \\
$\hat{\psi}^{(1)}(z_c)\sim const$
\\ \\
So we see the 1st KK state being more and more concentrated at the 2nd brane as
 $z_c$ decreases and $E$ increases. As $E\rightarrow\infty$ we find this
effect of
$\hat{\psi}^{(1)}(z_c)>>\hat{\psi}^{(1)}(0)$ becomes even more
pronounced. Thus as the second brane approaches the turn-around point in the warp factor, the mass of the previously 'ultralight' mode actually blows up, and this mode is concentrated on the second brane. We have a theory with only a single massless graviton. This is what we expect. As $z_c \rightarrow z_0 $ we see less and less of the geometry associated with the turn around in the warp factor. As it is the turn around that really differentiates between the physics corresponding to Minkowski and AdS branes, cutting off the space should result in a model similar to RS1.

\subsubsection{The Other KK Modes}

\noindent
As an interesting aside note that not only does the ultralight mode disappear in this limit, but infact we lose exactly half the KK modes when $z_c = z_0 $ because of the orbifold symmetry. 
$E$ is finite so
\begin{eqnarray}
&&\frac{C_1}{C_2}=-\frac{1}{Es_c}\rightarrow\infty
\nonumber\\
&&\Rightarrow C_2=0
\nonumber\\
&&\hat{\psi}^{(1)}(z)=C_1\cos^{\frac{5}{2}}(.)F_{\frac{5}{4}}
\end{eqnarray}

\noindent
This is consistent with the boundary condition at the first brane only if we
make $F_{\frac{5}{4}}$ finite, so
\begin{equation}
E=n(n+3),   n odd
\end{equation}

\noindent
We lose the states with $n$ even - that is those with wavefunctions that are
asymmetric about $z_0$, because of the orbifold condition at that point.

\subsubsection{What if $z_c < z_0$?}
\noindent
Let us briefly address the question of what happens when $z_c<z_0$. As earlier stated, we expect to approximately reproduce the results of RS because in this case we cut off the space before reaching the turn-around point in the warp factor. The second brane necessarily has negative tension given by
\begin{equation}
V_c = \frac{3}{L}\tanh(\frac{(y_c-y_0)}{L})
\end{equation}

\noindent
Let us suppose $y_c << y_0$. We find that we cannot consistently solve the boundary conditions for the first KK state if we assume that it is ultralight:
\begin{equation}
-1+\frac{3\epsilon_0}{E} + \frac{3\epsilon_0^2}{E} + ...= -1+\frac{3\epsilon_c}{E} + \frac{3\epsilon_c^2}{E} + ...
\end{equation}
\noindent
this is not possible for $\epsilon_0 \neq \epsilon_c$
\\ \\
On the other hand, looking at the zero mode we find that it is still normalizable:
\begin{eqnarray}
&&\hat{\psi}^{(0)}(z) = \frac{A^2}{\cos((|z|-z_0)/\hat{L})^3}
\nonumber\\
&&A^2= \frac{1}{\hat{L}(\frac{\sin(z_0/\hat{L})}{\cos(z_0/\hat{L})^2} - \frac{\sin((z_0 - z_c)/\hat{L})}{\cos((z_0-z_c)/\hat{L})^2})}
\end{eqnarray}

\noindent
As we always have $|z-z_0|<z_0$ we find $\hat{\psi}^{(0)}(z)$ is concentrated on the first brane and falls off towards the second brane. This is just as in the RS scenario. We reiterate that it is the turn-around in the warp factor that leads to the new and exciting results found in the Karch-Randall scenario - if the second brane cuts off the physics before we reach $z_0$ we won't see an ultralight mode.

\section{Implications}
\subsection{The Newton Potential}
\noindent
Now that we have investigated where the two gravtion modes are concentrated, we can finally find the form of the Newton potential between two unit masses on the first brane. We expect that both gravitons will contribute different amounts to the potential depending on the position of the second brane. For instance, because the ultralight mode becomes heavy as $z\rightarrow z_0 $ we should find that only the massless mode contributes in that case. 
\\ \\
If the masses are separated by distance $r$, then
\begin{eqnarray}
V(r)&=&\sum_{n}\frac{1}{r}e^{-mr}|\hat{\psi}^{(n)}(0)|^2\frac{1}{M^3}
\nonumber\\
&\approx& \frac{1}{r}\frac{1}{M^3}(|\hat{\psi}^{(0)}(0)|^2 + e^{-m_1r}|\hat{\psi}^{(1)}(0)|^2) + \int_2^{\infty}dn\frac{e^{-m_nr}}{M^3r}|\hat{\psi}^{(n)}(0)|^2
\end{eqnarray}

\noindent
For $r<<1/m_1$ the contribution of the first two gravitons is roughly
\begin{eqnarray}
&&\frac{1}{r}\frac{1}{M^3}|\hat{\psi}^{(0)}(0)|^2 \equiv  \frac{1}{M_{pl0}^2}\frac{1}{r}
\nonumber\\
&&\frac{1}{r}\frac{1}{M^3}|\hat{\psi}^{(1)}(0)|^2 \equiv  \frac{1}{M_{pl1}^2}\frac{1}{r}
\end{eqnarray}

\noindent
For instance, in the symmetric configuration,
\\ \\
$|\hat{\psi}^{(0)}(0)|^2 = \frac{M^3}{M_4^2} \approx |\hat{\psi}^{(1)}(0)|^2$
\\ \\
so that we get
\\ \\
$\frac{1}{r}\frac{1}{M_4^2} \equiv \frac{1}{M_{pl0}^2}\frac{1}{r} \equiv \frac{1}{M_{pl1}^2}\frac{1}{r}$
\\ \\

Here $M_4$ is the 4D scale obtained from dimensionally reducing the action.
We see that for ultralarge distances $r>>1/m_1$ the ultralight mode decouples. For nonsymmetric positionings of the two branes, $\hat{\psi}^{(0)}$ and $\hat{\psi}^{(1)}$ contribute different amounts to $M_{pl0}$ and $M_{pl1}$. For instance, as $y_c \rightarrow\infty$ only $\hat{\psi}^{(1)}$ provides a significant contribution ($M_{pl0}\gg M_{pl1}$), whereas when $y_c \rightarrow y_0$, $\hat{\psi}^{(0)}$ dominates instead.

\subsection{The Hierarchy Problem}
\noindent
One of the original motivations for the RS model was that it solved the hierarchy problem. We can ask whether the same is true of our model. 
As in Randall-Sundrum, physical masses are rescaled with respect to the fundamental mass parameter $m_0$ by
\begin{equation}
m=e^{A(y)}m_0
\end{equation}
If we put the SM fields on the second brane and use some form of stabilization mechanism to fix its position, we will need $y_c<2y_0$ for $e^{A(y_c)}<1$ as:
\\ \\
$e^{A(y)}=\frac{\cosh(\frac{|y|-y_0}{L})}{\cosh(\frac{y_0}{L})}$
\\ \\
However we have seen that if $y_c << 2y_0$ then only $\hat{\psi}^{(0)}$ mediates 4D gravity and we lose the bigravity characteristic of our model.

\subsection{Origins of the 2 Gravitons}
\noindent
In the bigravity model we have 2 gravitons combining in such a way that we see effective 4D gravity. The idea that we have 2 gravitons in the theory may seem somewhat mysterious. To better understand the origins of this scenario let us consider the symmetric configuration once more, but with both branes placed at an infinite separation from one another \cite{multi}.
\\ \\
We then have two independent branes with delta-function potentials, each supporting its own zero mode. When the branes are brought to a finite separation, these two modes mix, and are no longer degenerate in mass.
\\ \\
Earlier we obtained the form of the graviton wave function in the KR model which we will call $\hat{\psi}^{(1)}_{\infty}$. The graviton is localized on the brane at $0$ and the second brane is at $\infty$. We can obtain the form of the equivalent wavefunction $\hat{\psi}'^{(1)}_{\infty}$ which is instead centred on a second brane at $z_c$ (while assuming the first brane is at $-\infty$). We have:
\begin{equation}
\hat{\psi}'^{(1)}_{\infty}(z) \equiv \hat{\psi}^{(1)}_{\infty}(|z_c-|z||)
\end{equation}

\noindent
We should find that by taking linear combinations of these two functions, we end up with $\hat{\psi}^{(0)}(z)$ and $\hat{\psi}^{(1)}(z)$:
\begin{eqnarray}
&&\hat{\psi}^{(0)}(z) = \alpha\hat{\psi}'^{(1)}_{\infty}+\beta\hat{\psi}^{(1)}_{\infty}
\nonumber\\
&&\hat{\psi}^{(1)}(z) = \gamma\hat{\psi}'^{(1)}_{\infty}+\delta\hat{\psi}^{(1)}_{\infty}
\end{eqnarray}

\noindent
Now
\\ \\
$\hat{\psi}^{(1)}_{\infty} = C_2 \cos^{\frac{5}{2}}((|z|-z_0)/\hat{L})\Biggr[\frac{|F_\frac{7}{4}|}{|F_{\frac{5}{4}}|}F_{\frac{5}{4}} - \sin((|z|-z_0)/\hat{L})F_{\frac{7}{4}}\Biggr]$
\\ \\
and in the symmetric comfiguration when $z_c=2z_0$:
\begin{eqnarray}
\hat{\psi}'^{(1)}_{\infty}(z) &&\equiv \hat{\psi}^{(1)}_{\infty}(|z_c-|z||)
\nonumber\\
&&= C_2 \cos^{\frac{5}{2}}((|z|-z_0)/\hat{L})\Biggr[\frac{|F_\frac{7}{4}|}{|F_{\frac{5}{4}}|}F_{\frac{5}{4}} + \sin((|z|-z_0)/\hat{L})F_{\frac{7}{4}}\Biggr]
\end{eqnarray}

\noindent
Thus we find for the symmetric configuration that
\begin{eqnarray}
\frac{\hat{\psi}^{(1)}_{\infty} - \hat{\psi}'^{(1)}_{\infty}}{2}
&=& - C_2 \cos^{\frac{5}{2}}((|z|-z_0)/\hat{L})\sin((|z|-z_0)/\hat{L})F_{\frac{7}{4}}
\nonumber\\
&\equiv& \hat{\psi}^{(1)}(z)
\end{eqnarray}

\noindent
What about the orthogonal combination?
\begin{eqnarray}
\frac{\hat{\psi}^{(1)}_{\infty} + \hat{\psi}'^{(1)}_{\infty}}{2}
&=& - C_2 \cos^{\frac{5}{2}}((|z|-z_0)/\hat{L})\frac{|F_{\frac{7}{4}}|}{|F_{\frac{5}{4}}|}F_{\frac{5}{4}}
\nonumber\\
&\equiv& \hat{\psi}^{(0)}(z)
\end{eqnarray}

\noindent
This is not the form of $\hat{\psi}^{(0)}$ we know and love. However if we impose the boundary conditions on this new wavefunction, we find that it must be massless:
\\ \\
$\frac{d\psi}{dz}=\Biggr(-\frac{5}{2\hat{L}}\tan(.)+\frac{4}{\hat{L}}\sin(.)\cos(.)(1-\frac{E}{4})\frac{F_{\frac{9}{4}}}{F_{\frac{5}{4}}}\Biggr)\psi$
\\ \\
if $(m\hat{L})^2=0$ then $\frac{F_{\frac{9}{4}}}{F_{\frac{5}{4}}} = \cos^{-2}((|z|-z_0)/\hat{L})$ and this satisifies the boundary condition:
\\ \\
$\frac{d\psi}{dz} = \frac{3}{2\hat{L}}\tan(.)\psi$
\\ \\
Now once we substitute $E=0$ the new wavefunction reduces to the familiar form of $\hat{\psi}^{(0)}$:

\begin{equation}
\frac{\hat{\psi}^{(1)}_{\infty} + \hat{\psi}'^{(1)}_{\infty}}{2} \sim \cos^{\frac{-3}{2}}((|z|-z_0)/\hat{L}) \equiv\hat{\psi}^{(0)}(z)
\end{equation}

\noindent
Thus taking linear combinations of the KR ultralight modes located on each brane indeed gives us the two gravitons of the bigravity model. We can see from a different perspective, that the existence of the two gravitons themselves is a result of having two branes on either side of, and far from the turnaround point in the warp factor.

\subsection{Minkowski Limit}
\noindent
Let us now check the $\Lambda_4\rightarrow 0$ limit for our models. This is equivalent to taking $\frac{y_0}{L}\rightarrow\infty$. Is this a valid limit?
\\ \\
We know that $M_{pl}$ is a function of $\hat{\psi}^{(0)}(0)$ and $\hat{\psi}^{(1)}(0)$ (if we're on the first brane), and we want it to be finite. From dimensionally reducing the 5D action we find the 4D scale to be:
\\ \\
$M_4 = \frac{M^3L}{\cosh^2(\frac{y_0}{L})}(\frac{y_c}{L}+\sinh(\frac{y_c}{L})\cosh(\frac{y_c-2y_0}{L}))$
\\ \\
This is easily seen to be finite for $y_c<y_0$, $y_c=y_0$ and $y_c=2y_0$, even as we send $y_0\rightarrow\infty$.
\\ \\
Now in the KR model, $M_4$ is infinite, but all the contribution to $\frac{1}{M_{pl}^2} $ comes from $\hat{\psi}^{(1)}(0)^2$, and this is finite, even in the limit $y_0\rightarrow\infty$. Thus this is the correct limit to take.
\\ \\
Let's look at the case $y_c>y_0$ in the limit $y_0\rightarrow\infty$ for $y_c=y_0$, $y_c=2y_0$ and $y_c>>2y_0$. Consider

\begin{eqnarray}
\hat{\psi}^{(0)}(z)^2 &&= \frac{A^2}{\cos^3(\frac{|z|-z_0}{\hat{L}})}
\nonumber\\
&&\equiv
\frac{\cosh^2(\frac{y_0}{L})}{L(\frac{y_c}{L}+\sinh(\frac{y_c}{L})\cosh(\frac{y_c-2y_0}{L}))}\frac{\cosh^3(\frac{y_c-2y_0}{L})}{\cosh^3(\frac{y_0}{L})}
\end{eqnarray}

\noindent
For finite $|y|<y_0$:
\begin{equation}
\psi^{(0)}(y)^2\sim \frac{1}{L}e^{\frac{2y_0}{L}}e^{-\frac{y_c}{L}}e^{-\frac{|y_c-2y_0|}{L}}e^{-\frac{3|y|}{L}}
\end{equation}

Thus for $y_c=2y_0$, then $\psi^{(0)}(y)^2\sim\frac{1}{2L}e^{-\frac{3|y|}{L}}$
\\ \\
for $y_c>>2y_0$, $\psi^{(0)}(y)^2\sim\frac{1}{L}e^{-\frac{3|y|}{L}}e^{\frac{4y_0}{L}}e^{-\frac{2y_c}{L}}\rightarrow 0$ as before
\\ \\
and for $y_c=y_0$, $\psi^{(0)}(y)^2\sim\frac{1}{L}e^{-\frac{3|y|}{L}}$
\\ \\
Now let us check the behaviour of $\psi^{(1)}$ in this limit. Its mass goes to zero as $e^{-\frac{2y_0}{L}}$ for $y_c\geq 2y_0$.
\\ \\
For $y_c=y_0$ we found the mass of $\psi^{(1)}$ blows up like $\coth (\frac{y_c-y_0}{L})$ and this does not change as $y_c=y_0\rightarrow\infty$, so this mode is irrelevant.
\\ \\
However for $y_c=2y_0$,
\begin{eqnarray}
&&\hat{\psi}^{(1)}(z)=-C_2\cos^{-\frac{3}{2}}(\frac{z-z_0}{\hat{L}})|F_{\frac{7}{4}}|\sin (.)(1+\frac{\cos^2(.)}{2}+ ...)
\nonumber\\
&&\Rightarrow\psi^{(1)}(y)=-C_2\cosh^{\frac{3}{2}}(\frac{y-y_0}{L})|F_{\frac{7}{4}}|\tanh (.)(1+\frac{\cosh^{-2}(.)}{2}+ ...)
\end{eqnarray}

\noindent
So $\psi^{(1)}(y)^2\sim e^{-\frac{3|y|}{L}}$, and it should have the same normalization as $\psi^{(0)}(y)$.
\\ \\
For $y_c\rightarrow\infty$ we found that $\psi^{(1)}$, unlike $\psi^{(0)}$, is normalizable, and

\begin{eqnarray}
&&\hat{\psi}^{(1)}(z)=C_2\cos^{-\frac{3}{2}}(\frac{z-z_0}{\hat{L}})|F_{\frac{7}{4}}|\Biggr[\Biggr(1+\frac{\cos^2(.)}{4}E\Biggr) - \sin (.)\Biggr(1+\frac{\cos^2(.)}{2}(1+\frac{1}{2}E)\Biggr)\Biggr]
\nonumber\\
&&\sin(\frac{z-z_0}{\hat{L}})\equiv\tanh(\frac{y-y_0}{L})\sim -1
\end{eqnarray}

\noindent
So $\hat{\psi}^{(1)}(z)\sim\cosh^{\frac{3}{2}}(\frac{y_0-y}{L})\sim e^{-\frac{3}{2L}z}$
\\ \\
Thus we find that we obtain the RS model with $\psi^{(0)}$ and $\psi^{(1)}$ combining to produce the massless RS graviton:
\\ \\
For
\begin{eqnarray}
y_c=y_0 &\Rightarrow&  \psi^{(0)}\rightarrow\psi^{(0)}_{RS}
\nonumber\\
y_c=2y_0 &\Rightarrow& \psi^{(0)}, \psi^{(1)}\rightarrow\psi^{(0)}_{RS}
\nonumber\\
y_c>>2y_0 &\Rightarrow& \psi^{(1)}\rightarrow\psi^{(0)}_{RS}
\end{eqnarray}

\section{Conclusion}
\noindent
We find many interesting conclusions are to be drawn from studying the effects
of having 2 gravitons. While for a range of positions $z_c$ of the 2nd
brane they both have an effect (with each being concentrated on a different
brane, apart from for the symmetric configuration); we see that we will
inevitably lose one of these gravitons in certain limits.
\\ \\
When we send the 2nd brane off to infinity, thus ending up with only one brane, we lose the massless mode which then fails to be normalizable. However if we
bring the 2nd brane too close to the first - ie when it approaches the
turn-around in the warpfactor - the mass of the 1st KK state blows up, and it
can no longer fulfil its role as the 2nd graviton.
\\ \\
The interplay of the two gravitons should be open to a holographic interpretation \cite{hol} and this provides an interesting possibility for future investigation.

\section{Acknowledgements}
\noindent
I would especially like to thank Lisa Randall for her collaboration on this work. I would also like to thank Andreas Karch, Shinji Mukohyama and Matt Schwartz for many useful discussions.


\begin{thebibliography} {kr}
\bibitem{kr}
A ~Karch and ~L ~Randall,  {\it Locally localized gravity}, hep-th/0011156
\bibitem{big}
I ~Kogan, ~S ~Mouslopoulos,  and ~A ~Papazoglou {\it A new bigravity model with exclusively positive branes}, hep-th/0011141
\bibitem{rs}
L ~Randall and ~R ~Sundrum,  {\it An alternative to compactification}, hep-th/9906064
\bibitem{hol}
R ~Bousso and ~L ~Randall,  {\it Holographic domains of Anti de-Sitter space}, hep-th/0112080
\bibitem{m}
M ~Schwartz,  {\it The emergence of localized gravity}, hep-th/0011177
\bibitem{multi}
I ~Kogan, ~S ~Mouslopoulos, ~A ~Papazoglou and ~G ~Ross,  {\it Multi-localization in multi-brane worlds}, hep-th/0107307
\bibitem{0}
O ~DeWolfe, ~D ~Freedman, ~S ~Gubser and ~A ~Karch,  {\it Modeling the fifth dimension with scalars and gravity}, hep-th/9909134
\bibitem{mim}
A ~Miemiec,  {\it A power law for the lowest eigenvalue in localized massive gravity}, hep-th/0011160
\bibitem{kal}
N ~Kaloper,  {\it Bent domain walls as brane-worlds}, hep-th/9905210
\bibitem{thick}
C ~Csaki, ~J ~Erlich, ~T ~Hollowood and ~Y ~Shirman,  {\it Universal aspects of gravity localized on thick branes}, hep-th/0001033
\bibitem{por}
M ~Porrati,  {\it No Van Dam-Veltman-Zakharov discontinuity in Ads space}, hep-th/0011152

\end{thebibliography}
\end{document}